\documentclass[11pt]{article}
\usepackage{amsmath,amsthm,amsfonts,amssymb,amscd}
\usepackage{multirow,booktabs}
\usepackage[table]{xcolor}
\usepackage[export]{adjustbox}
\usepackage{fullpage}
\usepackage{lastpage}
\usepackage{enumitem}
\usepackage{fancyhdr}
\usepackage{mathrsfs}
\usepackage{wrapfig}
 \usepackage{relsize}
\usepackage{setspace}
\usepackage{calc}
\usepackage{multicol}
\usepackage{hyperref}
\usepackage{cancel}
\usepackage[retainorgcmds]{IEEEtrantools}
\usepackage[margin=3cm]{geometry}
\usepackage{amsmath}

\usepackage{empheq}
\usepackage{framed}
\usepackage[most]{tcolorbox}
\usepackage{xcolor}
\usepackage{braket}
\colorlet{shadecolor}{orange!15}
\theoremstyle{definition}

\usepackage{pifont}
\usepackage{bbm}

\definecolor{pansypurple}{rgb}{0.47, 0.09, 0.29}
\definecolor{patriarch}{rgb}{0.5, 0.0, 0.5}
\definecolor{carmine}{rgb}{0.59, 0.0, 0.09}
\definecolor{blueflag}{rgb}{0.2, 0.2, 0.6}
\definecolor{violet(ryb)}{rgb}{0.53, 0.0, 0.69}
\definecolor{operamauve}{rgb}{0.72, 0.52, 0.65}
\definecolor{olive}{rgb}{0.42, 0.56, 0.14}
\definecolor{mulberry}{rgb}{0.77, 0.29, 0.55}
\definecolor{prettyyellow}{rgb}{0.91, 0.84, 0.42}

\usepackage{cite}

\DeclareMathOperator{\bZ}{\mathbb{Z}}

\DeclareMathOperator{\bR}{\mathbb{R}}
\DeclareMathOperator{\bQ}{\mathbb{Q}}

\newcommand{\be}{\begin{equation}}
\newcommand{\ee}{\end{equation}}

\begin{document}
\setcounter{section}{0}
\title{}
\thispagestyle{empty}
	\fontsize{12pt}{20pt}	
	\vspace{13mm}  

 \color{black}
	\begin{center}
		{\huge  Non-Invertible SO(2) Symmetry of 4d Maxwell from Continuous Gaugings
  }
		\\[13mm]
		{\large  Elise Paznokas$^{\text{\ding{95}}}$}	
				\bigskip
				
				{\it
						 $^{\text{\ding{95}}}$Physique Th\'eorique et Math\'ematique and International Solvay Institutes\\
Universit\'e Libre de Bruxelles, C.P. 231, 1050 Brussels, Belgium  \\
\bigskip 
{elise.paznokas at ulb.be}

					}
		
	\end{center}

\begin{abstract}
    We describe the self-duality symmetries for 4d Maxwell theory at any value of the coupling $\tau$ via topological manipulations that include gauging continuous symmetries with flat connections. Moreover, we demonstrate that the $SL(2,\bZ)$ duality of Maxwell can be realized by trivial gauging operations. Using  a non-compact symmetry topological field theory (symTFT) to encode continuous global symmetries of the boundary theory, we reproduce the symTFT for Maxwell and find within this framework condensation defects that implement the non-invertible $SO(2)$ self-duality symmetry. These defects are systematically constructed by higher gauging subsets of the bulk $\bR\times \bR$ symmetry with appropriate discrete torsion. 
\end{abstract}

\newpage
\pagenumbering{arabic}
\setcounter{page}{1}
\setcounter{footnote}{0}
\renewcommand{\thefootnote}{\arabic{footnote}}

{\renewcommand{\baselinestretch}{.88} \parskip=0pt
	\setcounter{tocdepth}{2}
	\tableofcontents}


\section{Introduction}
The seminal paper of \cite{Gaiotto:2014kfa}, offering a new, more broad perspective of symmetries as topological operators of any codimension, has ushered in a wave of research into so-called generalized symmetries. This modern langauge has allowed for a more unified approach to describe $p$-form symmetries, discrete symmetries, non-invertible symmetries, and self-duality symmetries. For reviews of this ever-expanding field, we refer the reader to \cite{Cordova:2022ruw,Brennan:2023mmt,Bhardwaj:2023kri,Iqbal:2024pee} and references therein. 

More recently, this methodology has been encoded into the symmetry topological field theory (symTFT) framework \cite{Gaiotto:2020iye,Freed:2022qnc,Apruzzi:2021nmk}, where one uses a TFT in $d+1$ dimensions to describe the symmetry structure of a $d$-dimensional QFT. This paradigm has proved quite useful, and has been applied to the study of non-invertible symmetries \cite{Burbano:2021loy,Kaidi:2022cpf,Bashmakov:2022uek,Antinucci:2022vyk,Bhardwaj:2023ayw}, anomalies \cite{Kaidi:2023maf, Antinucci:2023ezl,Cordova:2023bja,Putrov:2024uor}, theories of manifolds with boundary \cite{Bhardwaj:2024igy,GarciaEtxebarria:2024jfv}, to name a few. In a variation of the traditional setup, in which the bulk TFT utilizes compact-valued gauge fields and can only describe the discrete sub-sector of the boundary symmetries, the works of \cite{Antinucci:2024zjp,Brennan:2024fgj} describe how by using non-compact fields in the bulk TFT, one can reproduce continuous symmetries for the $d$-dimensional boundary.\footnote{Theories with continuous symmetries have also been described via a non-topological $d+1$ bulk in \cite{Apruzzi:2024htg}. Additionally, non-Abelian continuous symmetries have been reproduced by using a non-Abelian $d+1$-dimensional BF theory as in \cite{Bonetti:2024cjk}.} This non-compact symTFT has been used for example in \cite{Arbalestrier:2024oqg,Argurio:2024ewp,Antinucci:2024bcm}. 

In this note, highly motivated by \cite{Argurio:2024ewp} where the authors study the self-duality symmetries of the 2d compact boson, we describe the non-invertible self-duality symmetries of 4d free Maxwell theory from a bulk perspective. In particular, we wish to use a non-compact TFT as the bulk 5d theory, which will allow us to not only reproduce the $U(1) \times U(1)$ 1-form symmetries of the 4d theory, but also to find the defects implementing a non-invertible $SO(2)$ 0-form self-duality symmetry.\footnote{For more on continuous non-invertible symmetries described from a defect perspective see \cite{Bhardwaj:2022yxj}.} 

Maxwell theory has long been known to enjoy self-duality symmetries.\footnote{Topological duality interfaces have also been known to be constructed via a half-space gauging procedure, see for example \cite{Kaidi:2022cpf}. By performing gauging manipulations on half of the spacetime on which the theory is defined, and combining with a duality transformation such that it is equivalent to the original theory, the interface separating the two halves is a topological operator for the duality symmetry.} For example, in \cite{Choi:2021kmx, Kan:2024fuu} the authors compose discrete gaugings of $\bZ_N \times \bZ_M \subset U(1) \times U(1)$ with the $SL(2,\bZ)$ duality to describe self-duality symmetries for specific values of the coupling $\tau$. In \cite{Niro:2022ctq}, the authors approach the self-duality symmetry in terms of its generating defects, and find that for the defects to be topological the possible transformations correspond to $SO(2)$ rotations, matching the older result of \cite{Deser:1997mz}. However, armed with only discrete gaugings, they were unable to formulate the defects corresponding to irrational values of the coupling and rotation angles. More recently, the authors in \cite{Hasan:2024aow} come to the same conclusion that there exists a non-invertible $U(1)$ symmetry in Maxwell by noting that any value of the coupling $\tau$ is stabilized by $U(1) \subset SL(2,\bR)$. It is clear that defects corresponding to irrational rescalings of the coupling, can no longer be realized through discrete gaugings of the 4d symmetries. Instead, the authors assert in their paper that the defects associated with irrational rescalings are constructed by allowing for infinite series of discrete gaugings.

In the context of the present literature, we argue that the advantage of our construction, armed with non-compact gauge fields and gauging continuous symmetries with flat connections, is that it allows us to define the self-duality symmetry for any value of the coupling $\tau$. Moreover, we are able to neatly describe the defects that implement the entire $SO(2)$ non-invertible symmetry for any value of the rotation angle $\theta$. This work also provides one example of the equivalence of approaches of gauging continuous symmetries with flat connections and infinite series of discrete gaugings. It would be interesting to prove this equivalence more generally.

The structure of this note is as follows. In Section \ref{sec2}, we begin by reviewing the setup of our 5d non-compact symTFT, its topological operators, and the desired topological and physical boundary conditions. We also demonstrate that the 5d theory has an $SL(2,\bR)$ symmetry which becomes an $SO(2)$ symmetry in the 4d theory based on the chosen boundary conditions. In Section \ref{sec3}, we illustrate how by composing continuous gaugings with flat connections with the $SL(2,\bZ)$ duality, we can describe a self-duality symmetry for any value of $\tau$ in Maxwell theory. Additionally, we show that the $SL(2,\bZ)$ duality can be realized via trivial gaugings of different subgroups $\bZ_1 \subset U(1)^e \times U(1)^m$. Section \ref{sec4} concerns the construction of the condensation defects that generate the bulk 0-form symmetry, and the subset of these that generate the symmetries in 4d. For completeness, in Section \ref{sec5}, we compute the fusion products of these defects from both the summation (here often integrals) over operators and path integral perspectives, to highlight certain properties of the defects. Appendix \ref{app a}, provides a more detailed description of the relationship between gauging with flat connections and topological defects.

\section{The 5d SymTFT}\label{sec2}
We start by reviewing the construction of the 5d symTFT that correctly reproduces 4d Maxwell theory with $U(1) \times U(1)$ 1-form symmetry after slab compactification. Informed by \cite{Antinucci:2024zjp, Brennan:2024fgj}, the correct bulk theory to encode continuous symmetries is given by a 5d BF theory with $\bR$-valued 2-form gauge fields.
\begin{equation} \label{1}
    S_{5d} = \frac{i}{2\pi} \int_{M_5} b\,d\,c 
\end{equation}
Throughout, we will adopt the notation of using lower case letters for fields valued in $\bR$, and upper case letters for fields valued in $U(1)$. The gauge fields $b,c$ have the corresponding gauge transformations,\footnote{We emphasize that $\bR$-valued gauge fields do not have large gauge transformations, meaning all gauge transformations are well-defined exact forms.}
\begin{equation}
    b \to b + d \lambda_b \quad , \quad c \to c+d\lambda_c\,.
\end{equation}
The equations of motion, $db=0$ and $dc=0$, allow us to define the following gauge invariant topological surface operators, 
\begin{equation}
    \begin{split}
        U_x(\Gamma) &= e^{ix \int_\Gamma b} \qquad x \in \mathbb{R} \\
        V_y(\Gamma) &= e^{iy\int_\Gamma c} \qquad y \in \mathbb{R}\,.
    \end{split}
\end{equation}
where $\Gamma$ is a closed 2-dimensional surface in $M_5$. Note that there is no quantization condition on the charges due to the lack of large gauge transformations for $\bR$-valued gauge fields. Via the canonical quantization of the BF theory, it is straightforward to derive braiding relations between surface operators: 
\begin{equation}
    \mathcal{B}( U_x(\Gamma),V_y(\Gamma') ) = \exp{\left(2\pi i \,xy\, \text{Link}(\Gamma,\Gamma')\right)}\,.
\end{equation}
Here $\text{Link}(\Gamma,\Gamma')=1$ if $\Gamma$ intersects with the surface filling $\Gamma'$, and $\text{Link}(\Gamma,\Gamma')=0$ if there is no such intersection. The 5d bulk has $\bR \times \bR$ 2-form symmetry generated by the operators $U_x,V_y$. The nontrivial braiding between $U_x$ and $V_y$ signals a 't Hooft anomaly between the two factors of $\bR$. 

We take the bulk theory to be living on the slab geometry $M_5=M_4\times I$. In particular, this means that we must specify boundary conditions at either end of the interval, $\partial M_5 = M_4 \sqcup \Bar{M}_4$, for the total theory to be well-defined under the action variation principle. Here, $\Bar{M}_4$ is the manifold with orientation opposite to $M_4$. On one of the boundaries, we will define topological boundary conditions, which specify the global variant, and thus the symmetry content of the 4d theory. On the opposite boundary, we impose physical boundary conditions that encode the dynamics of the 4d theory. By choosing these boundary conditions in an informed manner, we will correctly reproduce the 4d Maxwell theory with $U(1)\times U(1)$ symmetry from the symTFT setup. 

\subsection{The Topological Boundary}
Firstly, we describe the topological boundary conditions that we fix on $M_4$. Notice that under the gauge transformation $b\to b+d\lambda_b$, the BF action in \eqref{1} changes by a boundary term,
\begin{equation}
    \delta S_{5d} = \frac{i}{2\pi} \int_{M_4} \lambda_b dc \,.
\end{equation}
Therefore, to ensure gauge invariance of the bulk + boundary action, we couple to topological edge modes, introducing the $U(1)$-valued 1-form gauge field $A$.
\begin{equation} \label{19}
    S_{4d,top.} = \frac{i}{2\pi} \int_{M_4} dA \wedge \left( \frac{1}{e} c - \frac{e \theta}{2\pi} b \right) - \frac{ie^2 \theta}{8\pi^2} \int_{M_4} b \wedge b 
\end{equation}
Notice that due to the degree of the forms involved we are able to introduce a discrete torsion term $\propto b \wedge b$. The field $A$ has the associated gauge transformation,
\begin{equation}
    A \to A + d\lambda_A - e \lambda_b \,,
\end{equation}
such that the combined action is gauge-invariant as desired. The above boundary condition is parametrized by the choice of constants $e$ and $\theta$, which can take any real values. After slab compactification, these will correspond to the electric charge and theta angle, respectively. 

The boundary equations of motion of $b\,, c$ both tell us, \label{88}
\begin{equation}
    b|_{M_4} = -\frac{1}{e} dA \,,
\end{equation}
and the sum over the edge mode fluxes, $\int dA \in 2\pi \bZ$, implies both 
\begin{equation} \label{3}
     \int_{\Gamma \in M_4} e\, b=2\pi \bZ \quad , \quad \int_{\Gamma' \in M_4} \left(\frac{1}{e} c - \frac{e\theta}{2\pi} b \right)\in 2\pi \mathbb{Z} \,.
\end{equation}
Due to this choice of boundary conditions, the surface operators with the following lattice of charges are able to trivially end on $M_4$,
\begin{equation} \label{8}
    \begin{aligned}
        U_{n\,e} \quad , \quad 
        U_{-\frac{e \theta m}{2\pi}} V_{m/e}  \qquad n,m \in \mathbb{Z}\,.
    \end{aligned}
\end{equation}
This set of operators forms a Lagrangian algebra \cite{Antinucci:2024zjp}.\footnote{In general, a Lagrangian algebra is defined as the maximal set of mutually transparent operators that generate an anomaly free symmetry whose gauging trivializes the theory \cite{Antinucci:2024bny, Bhardwaj:2024qiv}.} This Lagrangian algebra is more general, for example than that in \cite{Argurio:2024ewp}, with the inclusion of the operators with $\theta$ dependent charges, allowed due to the fact that in this case the linking is antisymmetric
\begin{equation} \label{braid}
    \mathcal{B}(U_{n}V_m(\Gamma), U_a V_b(\Gamma')) = e^{2\pi i (nb-ma)}\,,
\end{equation}
which ensures that the operators are mutually transparent \cite{Antinucci:2024bcm, Antinucci:2024bny}. 

The total operators modulo the boundary ending operators are those with charges $\in \bR/\bZ$ and therefore they generate the $U(1)^e \times U(1)^m$ 1-form symmetry present on the boundary. The line operators that are charged under this 1-form symmetry in 4d are the endlines of the boundary-ending surfaces in \eqref{8}. Thus, in the presence of non-trivial $\theta$, the magnetic charged operators are dressed by electric charges. This phenomenon is the well-known Witten effect \cite{WITTEN1979283}.

The limits $e \to0 ,\infty$ correspond to full Dirichlet boundary conditions for either $b$ or $c$ respectively, such that in that case the boundary theory has $\bR$ symmetry. In these limits, there is an absence of a sensible $b\wedge b$ term, meaning that after slab compactification there is no theta term. This is as expected as for a `non-compact Maxwell theory' with $\bR$ symmetry the theta term would be a total derivative and thus would vanish. All other possible choices of the topological boundary conditions can be obtained by an $SL(2,\bR)$ action on either the Dirichlet or the aforementioned `mixed' boundary conditions \cite{Antinucci:2022vyk, Hasan:2024aow}. 

\subsection{The Physical Boundary and Compactification}
On the opposite boundary $\Bar{M}_4$, we impose physical boundary conditions that encode the dynamics of the 4d theory. In principle, by the symTFT prescription, we are allowed to add any 4d QFT that has the same 1-form symmetries as specified by the topological boundary. Here we do so by choosing the following physical boundary term,
\begin{equation} \label{122}
    S_{4d,phys. } = -\frac{1}{4\pi} \int_{\Bar{M}_4}  c \wedge \star_{4d} \, c\,.
\end{equation}
which depends on the metric on the boundary.\footnote{Boundary conditions with analogous structure to those in \eqref{122} have been discussed in \cite{Maldacena:2001ss}, where such boundary conditions give rise to generalized Maxwell theories.  More recently, in \cite{Antinucci:2024bcm} the authors considered a bulk (gauged) TQFT with a single physical boundary similar to that here as a holographic dual to the low energy effective field theory of a theory with spontaneous symmetry breaking. See also \cite{Benini:2022hzx}.}
As the fields are $\mathbb{R}$-valued, we have the freedom to rescale the fields to set the a priori arbitrary prefactor here to one, a fact which is important for our interpretation of $e$ as the electric charge. The equations of motion on this boundary are $b |_{\Bar{M}_4}= -i \star_{4d} c\,$.

Upon performing slab compactification, we combine both boundary terms and evaluate them on-shell.  The 4d boundary theory becomes,
\begin{equation} \label{27}
    S_{4d} = S_{top.} - S_{phys.} = \frac{1}{4\pi e^2} \int_{M_4} dA \wedge \star dA + \frac{i\theta}{8\pi^2} \int_{M_4} dA \wedge dA\,,
\end{equation}
which is the action for 4d Maxwell theory with a theta term that has the periodic identification $\theta \cong \theta +2\pi$ due to the topology of the $U(1)$ gauge fields. 
We can specify the above theory via the complex coupling parameter,\footnote{Note that we can achieve the usual normalization of Maxwell theory (see for example \cite{Witten:1995gf}) by rescaling $e^2 \to e^2/4\pi$ in our choice of topological boundary conditions. However, for our purposes we will keep the normalization in \eqref{27} as it makes the discussion of the operator spectrum and other aspects less cluttered with factors of $\sqrt{2\pi}$.}
\begin{equation}
    \tau =\frac{i}{e^2}+  \frac{\theta}{2\pi} \,.
\end{equation}

\subsection{Symmetries of the SymTFT}\label{symmetries}
Having delineated the setup of the 5d bulk and its boundary conditions, we can also examine the 0-form symmetries of this system. A similar discussion for the 3d/2d case can be found in \cite{Argurio:2024ewp}, where the difference with respect to 5d/4d is that the symmetry of the 2d theory is discrete $\bZ_2 \times \bZ_2$, while here it is found to be continuous. 

In particular, the 0-form symmetries of the 5d bulk are the transformations which preserve the form of the bulk action in\eqref{1}. The most general linear transformation on the gauge fields is given by,
\begin{equation} \label{4}
    \begin{pmatrix}
        \Tilde{b} \\ \Tilde{c}
    \end{pmatrix} = \begin{pmatrix}
        A &B \\ C & D 
    \end{pmatrix} \begin{pmatrix}
        b \\ c
    \end{pmatrix}\,,
\end{equation}
where $A,B,C,D \in \bR$ are constants. Plugging this into the BF action,
\begin{equation}
    \begin{split}
        \Tilde{b}d\Tilde{c} = (Ab+Bc)d (Cb+Dc) = (AD-BC) \,bdc +AC\, bdb +BD\,cdc\,.
    \end{split}
\end{equation}
The last two terms are killed since for 2-forms $bdb =-bdb$, meaning that the only condition for invariance of the action is $AD-BC=1$. This allows us to conclude that the 5d theory has an $SL(2,\bR)$ 0-form symmetry. 

There should exist a subset of the above $SL(2,\bR)$ transformations which also preserves the physical boundary condition. Such transformations correspond to the (possibly non-invertible) symmetries of the 4d theory. That is, we wish to find the transformations that preserve the equation $b = -i \star c$. Plugging in the transformation from \eqref{4},
\begin{equation}
    Ab + B c =-i \star C b - i \star D c\,. 
\end{equation}
For the physical boundary condition to be preserved we must impose further restrictions,
\begin{equation}
    A = D \quad , \quad B=-C \,.
\end{equation}
Thus, the symmetries of the 4d are given by matrices of the form,
\begin{equation}
    \begin{pmatrix}
        A &B \\ -B & A 
    \end{pmatrix}  \quad, \quad  A^2 +B^2=1\,.
\end{equation}
This is a matrix representation of the special orthogonal group $SO(2,\bR)$. Interestingly, this symmetry is continuous and is isomorphic to the circle group $U(1)$. As emphasized in \cite{Hasan:2024aow}, this signals a reemergence of the symmetry $U(1)$ which exists in the classical theory. The classical symmetry is broken in the quantum regime, not by anomalies, but because its action results in improperly quantized electric and magnetic fluxes. We find that this $U(1)$ symmetry reappears in the quantum theory as a non-invertible symmetry. 

\section{Gauging and Self-Duality Symmetries in Maxwell} \label{sec3}
In this section, we will show that by generalizing the gauging procedure to include gauging continuous symmetries with flat connections, we are able to define a (possibly non-invertible) self-duality symmetry for any value of the coupling $\tau$. Similarly to that in \cite{Argurio:2024ewp}, we perform a two-step gauging procedure:
\begin{enumerate}
    \item Gauge the entire $U(1)^m$ magnetic symmetry with flat connections. 
    \item Gauge $\bZ \subset \bR^e$ with appropriate discrete torsion. 
\end{enumerate}

Let us demonstrate this procedure from the path integral perspective, starting with Maxwell theory with $\tau = \frac{i}{e^2}+ \frac{\theta}{2\pi}$.
\begin{equation}
    S= \frac{1}{4\pi e^2} \int dA \wedge \star dA + \frac{i\theta}{8\pi^2} \int dA \wedge dA\,.
\end{equation}
To gauge the $U(1)^m$ symmetry, we couple the associated conserved current $\star J =\frac{i}{2\pi} dA$ to a dynamical 2-form gauge field $\Phi$ that is valued in $U(1)$.
\begin{equation}
    S^{U(1)\; gauged} = S + \frac{i}{2\pi} \int dA \wedge \Phi - \frac{i\, e}{2\pi} \int b \wedge d\Phi 
\end{equation}
We have included the Lagrange multiplier $b \in\bR$ to enforce the flatness of $\Phi$. The prefactor $e$ of this term is fixed for convenience. As $b$ is a real-valued field, $\int d b =0$, and it does not impose any quantization conditions on $\Phi$. Path integrating over $\Phi$ produces the delta function $\delta(dA=e\, db)$, such that the integral over $A$ enforces this delta, leaving,
\begin{equation}
    S^{U(1)\; gauged} = \frac{1}{4\pi} \int db \wedge \star db + \frac{i\theta \,e^2}{8\pi^2} \int db \wedge db\,.
\end{equation}
However, $b$ is real-valued, meaning the theta term is just a total derivative and it cancels since the four-manifold does not have a boundary.
\begin{equation}
    S^{U(1)\; gauged} = \frac{1}{4\pi} \int db \wedge \star db
\end{equation}
This has completed step 1) of the gauging procedure leaving us with `non-compact Maxwell' with an $\bR$ 1-form symmetry. 

For the second step, we then gauge a discrete subgroup of the electric symmetry $\bZ^e \subset \bR^e$ with a choice of discrete torsion.
\begin{equation}
    S^{\bZ \odot U(1)\, gauged} = \frac{1}{4\pi} \int (db-c) \wedge \star (db-c) - \frac{i}{2\pi e^\prime } \int c \wedge d\Tilde{A} + \frac{i\theta^\prime}{8\pi^2} \int d\Tilde{A} \wedge d\Tilde{A}
\end{equation}
The 2-form gauge field $c$ is introduced to gauge the symmetry generated by the conserved current $\star J= \star db$. The field $\Tilde{A} \in U(1)$ is included to restrict the holonomies of $c$ to be discrete. The final term is a discrete torsion term that we can generically include. Performing the path integral over $c$ gives $\delta( \star (db-c)= \frac{i}{e^\prime}  d\Tilde{A})$ such that,
\begin{equation}
    S^{\bZ \odot U(1)\, gauged} = \frac{1}{4\pi(e^\prime)^2} \int d\Tilde{A} \wedge \star d\Tilde{A}+ \frac{i\theta^\prime}{8\pi^2} \int d\Tilde{A} \wedge  d\Tilde{A}\,.
\end{equation}
This concludes step 2). We see that we are left with Maxwell at new coupling,
\begin{equation}
    \tau^\prime =  \frac{i}{(e^\prime)^2}+ \frac{\theta^\prime}{2\pi} \,.
\end{equation}
We emphasize that the beauty of this two-step procedure is that it allows us to manipulate the coupling $\tau$ to any new $\tau'$, since there are no restrictions on how we can choose $e', \, \theta'$. These manipulations are topological ones, a point that we describe in more detail in Appendix \ref{app a}.

In order to define the self-duality symmetry for any value of $\tau$,\footnote{In contrast, by only allowing for gauging of discrete non-anomalous subsets of the 1-form symmetry, $\bZ_N \times \bZ_M \subset U(1) \times U(1)$ with $gcd(N,M)=1$, this means that the self-duality symmetry is only definable for a priori specific values of $\tau$. These specific values are the solutions to $\tau = \frac{a(\frac{M}{N})^2\tau +b}{c(\frac{M}{N})^2\tau +d}$.} we first implement the two-step gauging procedure, tailoring $e^\prime$ and $\theta^\prime$ in such a way that,
\begin{equation}
    \tau^\prime = \frac{d\tau-b}{-c\tau +a}\,.
\end{equation}
Then, by invoking the duality transformation $\begin{pmatrix}a&b\\ c & d \end{pmatrix} \in SL(2,\bZ)$, this sends $\tau^\prime \to \tau$ such that we are back at the original coupling. This self-duality symmetry is clearly non-invertible due to the fact that the gauging procedure involves changing the spectrum of genuine and twist operators.

\subsection{SL(2,Z) via Trivial Gauging}
In addition, we can also define the $SL(2,\bZ)$ transformations via trivial gauging operations. By trivial gauging, we mean gauging different subsets $\bZ_1\subset U(1)^e \times U(1)^m$ (with appropriate discrete torsion) such that we are only gauging the identity and the operation is invertible.\footnote{The authors in \cite{Argurio:2024ewp} find that the $T$ duality for the 2d compact boson can be defined from the trivial gauging of the $\bZ_1$ subset of the shift symmetry.} As $SL(2,\bZ)$ is generated by the $S$ and $T$ transformations that act on the coupling as $S: \tau \to -1/\tau$ and $T: \tau \to \tau+1$, we will focus on these in what follows. 

Firstly, the $T$ transformation can be realized by gauging a trivial subset of the magnetic symmetry, $\bZ_1 \subset U(1)^m$, with discrete torsion. We begin with Maxwell theory with coupling $\tau=\frac{i}{e^2}+\frac{\theta}{2\pi}$,
\begin{equation}
    S(\tau) = \frac{1}{4\pi e^2} \int_{M_4} dA \wedge \star dA + \frac{i\theta}{8\pi^2} \int_{M_4} dA \wedge dA\,.
\end{equation}
The gauging is done by coupling the conserved current $\star J =\frac{i}{2\pi} dA$ to the 2-form $U(1)$-valued gauge field $\Phi$, along with Lagrange multiplier and discrete torsion terms.
\begin{equation}
    S^{\bZ_1^m \; gauged}(\tau) = S(\tau) - \frac{1}{2\pi}\int dA \, \Phi +\frac{i}{2\pi}\int d\Psi\,\Phi + \frac{i}{4\pi}\int \Phi \wedge \Phi\,.
\end{equation}
Integrating over the field $\Psi$ enforces that $\Phi$ is quantum-mechanically equivalent to a $\bZ_1$ gauge field, and thus the gauging is trivial. Completing the square and using Gaussian integration, the path integral over $\Phi$ yields,
\begin{equation}
    S^{\bZ_1^m \; gauged}(\tau) = S(\tau) +\frac{i}{4\pi}\int dA \wedge dA 
\end{equation} 
which is equivalent to $S(\tau+1)$, reproducing the action of $T\in SL(2,\bZ)$.

Moving on to the $S$ transformation, this can be achieved by trivially gauging the electric symmetry, $\bZ_1\subset U(1)^e$. Coupling the conserved current $\star J= \frac{1}{2\pi e^2}\star dA$ to the gauge field $\Phi \in U(1)$ in a gauge invariant manner,
\begin{equation}
    S^{\bZ_1^e \; gauged}(\tau) =\frac{1}{4\pi e^2}\int (dA- \Phi)\wedge\star (dA- \Phi) + \frac{i\theta}{8\pi^2}\int (dA- \Phi)\wedge (dA- \Phi) - \frac{i}{2\pi}\int d\Tilde{A} \Phi\,.
\end{equation}
We have introduced $\Tilde{A}$ which acts as a Lagrange multiplier, and whose sum over fluxes imposes $\int\Phi \in 2\pi \bZ$. The path integral over $\Phi$ produces the delta function 
\begin{equation}
    \delta\left( d\Tilde{A}= \frac{i}{e^2}\star (d A-\Phi) - \frac{\theta}{2\pi} (dA-\Phi) \right)\,.
\end{equation}
Integrating over $A$ then yields,
\begin{equation}
   S^{\bZ_1^e \; gauged}(\tau) = \frac{1}{4\pi e^2(\frac{1}{e^4}+\frac{\theta^2}{4\pi^2})}\int d\Tilde{A} \wedge \star d \Tilde{A} - \frac{i\theta}{8\pi^2 (\frac{1}{e^4}+\frac{\theta^2}{4\pi^2})} \int  d\Tilde{A} \wedge d \Tilde{A}
\end{equation}
such that we see this trivial gauging operation served to swap the field strength $F$ with its dual $\Tilde{F}= d\Tilde{A}= \frac{i}{e^2} \star dA - \frac{\theta}{2\pi} dA$ and the coupling $\tau \to -\frac{1}{\tau}$. This is exactly the $S$ transformation, $S^{\bZ_1^e \; gauged}(\tau) = S(-1/\tau)$. 

\section{The Condensation Defects} \label{sec4}
Thus far, we have discussed the surface operators $U,V$, which generate the $\bR \times \bR$ symmetry in the bulk. In addition to these, we can also define codimension-one defects via higher gauging, so-called condensation defects, which will generate the 0-form symmetries of the bulk TQFT. Such codimension-one defects act non-trivially on the surface operators. 

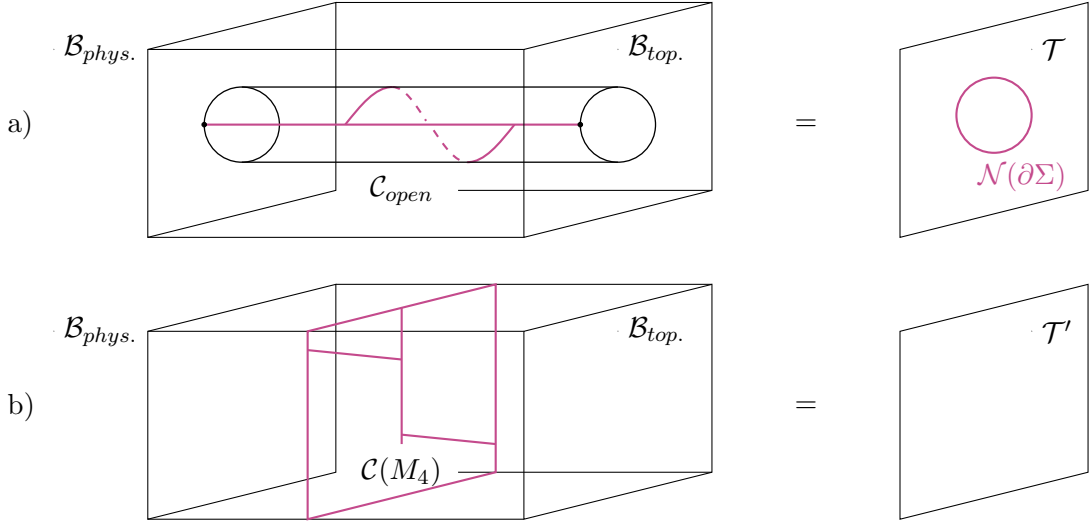
\begin{figure}[t]
\centering
    \scalebox{1}{
    \begin{tikzpicture}[scale=2.5]
            \filldraw(-.8,.6) circle (.01 pt) node[color=black, anchor=west]{a)};
            \filldraw(2.5,1) circle (.01 pt) node[color=black, anchor=west]{$\mathcal{B}_{top.}$};
            \filldraw(-.5,1) circle (.01 pt) node[color=black, anchor=west]{$\mathcal{B}_{phys.}$};
            \filldraw(4.7,1) circle (.01 pt) node[color=black, anchor=west]{$\mathcal{T}$};
		\draw (1.65,0.25) -- (3,0.25); \draw (1,1.25) -- (3,1.25); 
		 \draw[] (0,0) -- (1,0.25) -- (1,1.25) -- (0,1) -- cycle;	
		 \draw (0,1) -- (2,1); \draw (0,0) --(2,0);
		   \draw[line width=0.5, color=black,opacity=0.8] (0.7, 0.6) arc (0: 360: 0.2);

		\node at (1.35,0.25) {$\mathcal{C}_{open}$};

		\draw[line width=0.8, mulberry] (1.05,0.6) sin (1.3,0.8);
		\draw[line width=0.8, mulberry, dashed] (1.3,0.8) cos (1.48,0.62);
		\draw[line width=0.8, mulberry,dashed] (1.52,0.57) sin (1.7,0.4);
		\draw[line width=0.8, mulberry] (1.7,0.4) cos (1.95,0.6);
		\draw[line width=0.8, color=mulberry] (2.3,0.6)--(0.3,0.6);
		
		\draw[fill=black] (0.3,0.6) circle (0.012);
		\draw[fill=black] (2.3,0.6) circle (0.012);

		\draw[] (2,0) -- (3,0.25) -- (3,1.25) -- (2,1) -- cycle;
   		 
		  \draw[line width=0.5, color=black] (0.5, 0.4)--(2.5,0.4);
		  \draw[line width=0.5, color=black] (0.5, 0.8)--(2.5,0.8);
		  \draw[line width=0.5, color=black] (2.7, 0.6) arc (0: 360: 0.2);

   \begin{scope}[shift={(4, 0.)}]

   		\node at (-0.5,0.6) {$=$};

		 \draw[] (0,0) -- (1,0.25) -- (1,1.25) -- (0,1) -- cycle;	
		 \draw[line width=0.8, color=mulberry,opacity=0.8] (0.7, 0.65) arc (0: 360: 0.2);  
		 \node at (0.65,0.32) { \color{mulberry}$\mathcal{N}(\partial \Sigma)$};

	\end{scope}

    \begin{scope}[shift={(0,-1.5)}]
    \filldraw(-.8,.6) circle (.01 pt) node[color=black, anchor=west]{b)};
            \filldraw(2.5,1) circle (.01 pt) node[color=black, anchor=west]{$\mathcal{B}_{top.}$};
            \filldraw(-.5,1) circle (.01 pt) node[color=black, anchor=west]{$\mathcal{B}_{phys.}$};
            \filldraw(4.7,1) circle (.01 pt) node[color=black, anchor=west]{$\mathcal{T}'$};
		\draw (1.65,0.25) -- (3,0.25); \draw (1,1.25) -- (3,1.25); 
		 \draw[] (0,0) -- (1,0.25) -- (1,1.25) -- (0,1) -- cycle;	
		 \draw (0,1) -- (2,1); \draw (0,0) --(2,0);

		\node at (1.35,0.25) {$\mathcal{C}(M_4)$};

		\draw[] (2,0) -- (3,0.25) -- (3,1.25) -- (2,1) -- cycle;

    \begin{scope}[shift={(.85, 0.)}]

		 \draw[line width=0.8,color=mulberry] (0,0) -- (1,0.25) -- (1,1.25) -- (0,1) -- cycle;
            \draw[line width=0.8,color=mulberry] (.5,.4) -- (.5,1.125);  
            \draw[line width=0.8,color=mulberry] (0,.9) -- (.5,.85);
            \draw[line width=0.8,color=mulberry] (.5,.45) -- (1,.4);
         \end{scope}
    
   \begin{scope}[shift={(4, 0.)}]

   		\node at (-0.5,0.6) {$=$};

		 \draw[] (0,0) -- (1,0.25) -- (1,1.25) -- (0,1) -- cycle;	

	\end{scope}
        
    \end{scope}
   
\end{tikzpicture}
}
\hspace{1cm}
    \caption{Condensation defects in the symTFT, and their different interpretations after slab compactification. In a), the open defect that upon compactification becomes a symmetry generator $\mathcal{N}$ of the 4d theory $\mathcal{T}$. In b), the defect defined on the manifold $M_4$ parallel to the boundary, such that pushing the defect to $B_{top.}$ generically produces a new topological boundary condition $\mathcal{B}_{top.}'$. After compactifying, this produces a new global variant of the 4d theory $\mathcal{T}'$. }
    \label{fig: defects inside slab}
\end{figure}

We emphasize that the placement of the condensation defects inside of $M_5$ will influence their interpretation in 4d after slab compactification. For example, if the defect is defined on a manifold parallel to the topological boundary, it can be pushed to the boundary to generically give rise to new topological boundary conditions. Alternatively, we can open the defects along non-trivial cycles, defining Dirichlet boundary conditions for the boundary ending operators to ensure gauge invariance. After slab compactification, these open defects become the symmetry generators of the 0-form symmetry in the 4d theory, see figure \ref{fig: defects inside slab}.

In this section we aim to accomplish the following:
\begin{enumerate}
    \item Identify the set of condensation defects that generate the $SL(2,\bR)$ 0-form symmetry of the bulk 5d. 
    \item Pick out a subset of these defects which are symmetries of the 4d theory, namely those that leave the 4d physical boundary condition invariant. 
    \item Consider open defects of the above type; these become generators of the $SO(2)$ non-invertible self-duality symmetry after slab compactification.
\end{enumerate}

\subsection{SL(2,R) Defects}
We first set out to identify the defects which are the generators of the bulk 0-form symmetry. In §\ref{symmetries}, we identified the 0-form symmetry of the 5d theory to be $SL(2,\bR)$. From the Iwasawa decomposition, we know that $SL(2,\bR)$ is generated by the matrices,
\begin{equation} \label{5}
    \begin{aligned}
        K_\vartheta &= \begin{pmatrix}
            cos\vartheta & -sin\vartheta \\sin\vartheta & cos\vartheta
        \end{pmatrix} \qquad &\vartheta &\in \bR\\
        T_\ell&= \begin{pmatrix}
            1 & 0 \\ \ell&1 \\
        \end{pmatrix} \qquad &\ell &\in \bR/\{0\}\\
        A_\lambda &= \begin{pmatrix}
            \lambda & 0 \\ 0 & \lambda^{-1}
        \end{pmatrix}\qquad &\lambda &\in \bR \,.
    \end{aligned} 
\end{equation}

To construct the condensation defects, we use Poincar\'e duality to rewrite the higher gauging as summations of the surface operators $U,V$ along non-trivial 2-cycles of the codimension-one manifold $\Sigma \subset M_5$.
\begin{equation}
    C(\Sigma, \mathcal{A}) = \frac{1}{\#}\sum_{\Gamma\in H_2 (\Sigma, \mathcal{A})} W(\Gamma) e^{2\pi i \alpha Link(\Gamma, \Gamma)}
\end{equation}
Schematically, this corresponds to gauging the 2-form symmetry $\mathcal{A}$ which is generated by the operator $W$ along with a choice of discrete torsion parametrized by $\alpha$. The discrete torsion phase is non-zero for two-cycles $\in H_2(\Sigma)$ that non-trivially link with each other, see for example \cite{Choi:2022zal} for more on discrete torsion in the case of discrete gaugings. In the following we will construct the condensation defects for the bulk $SL(2,\bR)$ (invertible) symmetry. For such invertible defects, we fix the normalization such that when acting with the defect on simple surface operators it returns simple surface operators with unit prefactor.\footnote{This is in contrast to the discrete case where normalization is fixed by $|H_2(\Sigma,\mathcal{A})|^{-1/2}$. For more on the construction, fusion, and properties of condensation defects in the discrete case, we refer the reader to \cite{Roumpedakis:2022aik, Shao:2023gho}.}

We are largely motivated by the work in \cite{Antinucci:2022vyk}, where the authors construct defects corresponding to the $SL(2,\bZ_N)$ 0-form symmetry of 4d $su(N)$ $\mathcal{N}=4$ super-Yang Mills via the symTFT setup,  to inform our ansatz for the construction of defects that correspond to $A,T,K$. Condensation defects made by gauging discrete symmetries in 4d Maxwell have also been constructed for example in \cite{Choi:2022zal, Choi2}.

\subsubsection{T Defect} \label{Tdefect}
Let us begin by determining the defect corresponding to the $T$ transformation, which acts on the vector of gauge fields $(b,c)^T$ with the matrix in \eqref{5}. The vector of charges $(n,m)^T$, carried by the respective surface operators, i.e. $U_n(\Gamma), \, V_m(\Gamma')$, transforms under the transpose matrix $T_\ell^T$,
\begin{equation} \label{14}
    T_\ell: (n,m) \to (n+\ell m,m)\,.
\end{equation}
 
To construct such a defect we higher gauge the $\mathbb{R}^e$ symmetry with discrete torsion.
\begin{equation}
    C^T(\Sigma) \propto \sum_{\Gamma \in H_2 (\Sigma, \bR)} U(\Gamma) e^{-2\pi i \ell^{-1} Link(\Gamma, \Gamma)}
\end{equation}
We change this formal sum into an integral over the $\bR$ charges of the operators living along the non-contractible 2-cycles of $\Sigma$. In this section, we will consider the situation where $\Sigma$ is closed, leaving the consideration of open defects for later. Following \cite{Antinucci:2022vyk}, we also take the four-manifold to be $\Sigma = T^2 \times S^2$.\footnote{The reasoning behind this choice is that we wish to choose the simplest manifold that will `encircle' the surface operators. Surrounding the surface operator by a tubular neighborhood gives the natural choice $T^2 \times S^2$.}
\begin{equation}
    C^T(T^2 \times S^2) = \ell^{-1} \int_{x,y \in \mathbb{R}} e^{-2\pi i \ell^{-1} xy} U_x(T^2) U_y(S^2) 
\end{equation}
Here we have decomposed $\Sigma=T^2 \times S^2$ into the generating 2-cycles $T^2$ and $S^2$. The phase factor comes directly from our choice of discrete torsion; there are no normal ordering phases here since $U$ does not braid non-trivially with itself. In what follows, we demonstrate that this is the correct choice of normalization, ensuring that the action of the defect on surface operators returns surface operators with prefactor one.

Let us consider acting on the operators $U_n(T^2) V_m(T^2)$. We will always choose by our convention to act on operators living on $T^2$, this will also inform how we resolve the junction of defects and pick up normal ordering phases.
\begin{equation}
    C^T(T^2 \times S^2)\, U_n(T^2) V_m(T^2)=  \ell^{-1}\int_{x,y \in \mathbb{R}} e^{-2\pi i\ell^{-1} xy} e^{2\pi i ym}  U_{x+n}(T^2) V_m(T^2) \,.
\end{equation}
Here we have unlinked $V_m$ with $U_y$, picking up a phase, and then fused like $U$ defects. The integral over $y$ produces the Dirac delta function, 
\begin{equation}
    \int_{-\infty}^\infty dy \,e^{2\pi i y \ell^{-1}(\ell m-x)} = \ell\; \delta(x=\ell m)\,,
\end{equation}
such that the integral over $x$ enforces the delta function and we are left with,
\begin{equation}
    C^T(T^2 \times S^2)\, U_n(T^2) V_m(T^2)=U_{n+\ell m}(T^2) V_m(T^2) \,.
\end{equation}
The action of the defect is as expected for the $T_\ell$ transformation.

\subsubsection{A Defect}
Moving on, we wish to describe the condensation defect that corresponds to the $A_\lambda$ transformation in \eqref{5}. The action on the vector of charges is,
\begin{equation} \label{6}
    A_\lambda: (n,m) \to (\lambda n, \lambda^{-1}m)\,.
\end{equation}

The associated condensation defect is constructed from gauging the entire $\bR \times \bR$ with discrete torsion parametrized by $\Lambda = \frac{\lambda +1}{2(\lambda - 1)}$.
\begin{equation}
    \begin{split}
        C^A(T^2\times S^2) & =(\Lambda^2 - \frac{1}{4})  \sum_{\Gamma, \Gamma' \in H_2(T^2\times S^2, \bR)} U(\Gamma) V(\Gamma')\;e^{2\pi i \Lambda Link(\Gamma, \Gamma')} \\
        & = (\Lambda^2 - \frac{1}{4}) \int_{x,y,w,z \in \bR} e^{2\pi i \Lambda (wz+xy)} U(wT^2 +x S^2) V(yT^2 + zS^2)
    \end{split}
\end{equation}
Resolving the junction of operators, we pick up a normal ordering phase which is given by the square root of the braiding in \eqref{braid}, see \cite{Roumpedakis:2022aik, Antinucci:2022vyk,Argurio:2024ewp}.
\begin{equation} \label{18}
    C^A(T^2\times S^2) = (\Lambda^2 - \frac{1}{4})  \int_{x,y,w,z \in \bR} e^{2\pi i \Lambda(wz+xy)}e^{\pi i (xy-wz)} U_w(T^2) V_y(T^2) U_x(S^2) V_z(S^2)
\end{equation}
Acting on the operators $U_n(T^2) V_m(T^2)$, we unlink and then fuse like operators to find, 
\begin{equation}
    \begin{split}
        C^A(T^2\times S^2) \,U_n(T^2) V_m(T^2) &= (\Lambda^2 - \frac{1}{4}) \int_{x,y,w,z \in \bR} e^{2\pi i \Lambda(wz+xy)}e^{\pi i (xy-wz)} e^{2\pi i (xm - zn ) } \\
    &\times U_{w+n}(T^2) V_{y+m}(T^2) \,.
    \end{split}
\end{equation}
Performing the integrals over $x,z$,
\begin{equation}
    \begin{split}
        \int_{\infty}^\infty dx \,e^{2\pi i x(\Lambda y + \frac{1}{2} y +m)} &= (\Lambda +1/2)^{-1} \,\delta\Big( y=-\frac{m}{\Lambda+1/2}\Big)\\
        \int_{\infty}^\infty dz \,e^{2\pi i z(\Lambda w - \frac{1}{2} w -n )} &= (\Lambda- 1/2)^{-1} \, \delta\Big(w=\frac{n}{\Lambda - 1/2}\Big)\\
    \end{split}
\end{equation}
such that 
\begin{equation}
    C^A(T^2\times S^2)\, U_n(T^2) V_m(T^2) = U_{n(1+(\Lambda-1/2)^{-1} )}(T^2) V_{m(1-(\Lambda+1/2)^{-1})}(T^2)\,.
\end{equation}
Using the definition, $\Lambda = \frac{\lambda +1}{2(\lambda - 1)}$, we can express the transformation in terms of $\lambda$,
\begin{equation}
    C^A(T^2\times S^2) \,U_n(T^2) V_m(T^2) =U_{n\lambda}(T^2) V_{m\lambda^{-1}}(T^2)
\end{equation}
which is exactly the rescaling we expected from \eqref{6}. 

Note that the condensation defect corresponding to charge conjugation $(n,m) \to -(n,m)$ is given by $C^A$ with $\lambda=-1$, $\Lambda =0$.

\subsubsection{K Defect} \label{sdefect}
Lastly, we come to the defect that implements the rotation by $\vartheta$ in \eqref{5}. The associated transformation on the vector of charges,\footnote{We again emphasize that this provides a more concrete realization of the rotational transformation matching that of equation 44 in \cite{Hasan:2024aow}. Our construction, allowing for continuous gauging with flat connections, means the relationship between gauging and irrational rescalings of the coupling no longer breaks down, and we can more neatly formulate an $SO(2)$ defect parametrized by any $\vartheta \in \bR$.}
\begin{equation} \label{7}
    K_{\vartheta} : (n,m) \to (n\cos\vartheta +m\sin \vartheta ,- n\sin \vartheta +m\cos \vartheta) \,.
\end{equation}

Here, without motivation from the literature on what symmetries to gauge to build such a defect, we instead reverse engineer the form of the defect starting from the desired transformation on the charges/gauge fields. To begin, we anticipate that the defect will require gauging the full $\bR^e \times \bR^m$ symmetry with appropriate discrete torsion. 
\begin{equation}
    C^K(T^2\times S^2)  \approx \int_{a,b,c,d \in \bR} e^{...} e^{\pi i (bc-ad)} U_a(T^2) V_c(T^2) U_b(S^2) V_d(S^2)
\end{equation}
Here $e^{...}$ represents the for now arbitrary discrete torsion phase, and $e^{\pi i (bc-ad)}$ is the required normal ordering phase, see comments around \eqref{18}. Once again, by acting on the operators $U_n(T^2) V_m(T^2)$ we pick up two additional linking phases,
\begin{equation}
    C^K(T^2\times S^2)\, U_n(T^2) V_m(T^2) \propto  \int_{a,b,c,d \in \bR} e^{...} e^{\pi i (bc-ad)} e^{2\pi i (bm-dn)} U_{a+n}(T^2) V_{c+m}(T^2) \,.
\end{equation}
From the desired transformation in \eqref{7}, we ask that the phase factors, once integrated over, produce delta functions that impose, 
\begin{equation}
    \begin{split}
        a&= n (\cos\vartheta -1) +m\sin \vartheta\\
        c &= m (\cos \vartheta -1) -n\sin \vartheta\,,
    \end{split}
\end{equation}
such that in terms of $n,m$ this becomes
\begin{equation}
    \begin{split}
        n &= - \frac{a}{2} + \frac{c\sin\vartheta}{2(\cos \vartheta -1)} \\
        m&=-\frac{a\sin\vartheta}{2(\cos \vartheta -1)} - \frac{c}{2}\,.
    \end{split}
\end{equation}
The appropriate delta functions will come from the following integrals over $b,d$,
\begin{equation}
    \begin{split}
        &\int_{-\infty}^\infty db \,\exp\left[2\pi i b (m+\frac{a\sin\vartheta}{2(\cos \vartheta -1)}+ \frac{c}{2})\right] \\
        &\int_{-\infty}^\infty dd \,\exp\left[2\pi i d (-n- \frac{a}{2} +\frac{c\sin\vartheta}{2(\cos \vartheta -1)})\right] \,.\\
    \end{split}
\end{equation}
This also implies that the correct normalization of the defect is $\frac{1}{2(1-\cos \vartheta)} $. 
Grouping the phase factors neatly into the condensation defect, 
\begin{equation} \label{11}
    C^K(T^2\times S^2) =\frac{1}{2(1-\cos \vartheta)}  \int_{a,b,c,d \in \bR} e^{\pi i (\frac{\sin \vartheta}{\cos\vartheta -1}) (ab+dc)} e^{\pi i (bc-ad)} U_a V_c(T^2) \,U_b V_d(S^2)\,.
\end{equation}
To summarize, we found that the condensation defect corresponding to $K_\vartheta$ is from higher gauging the $\bR^e\times \bR^m$ 2-form symmetry with discrete torsion $\frac{\sin \vartheta}{2(\cos\vartheta -1)}$ for both of the respective factors.

\subsection{Open Defects} \label{open defects}
As discussed at the beginning of this section, the invertible $SL(2,\bR)$ symmetry in the 5d bulk is generated by the $C^T$, $C^A$, and $C^K$ defects defined on closed manifolds. (The invertibility of these defects is verified in the next section via fusion arguments). In §\ref{symmetries}, we determined that only the $K_{\vartheta}$ transformations leave the physical boundary conditions invariant. Thus, only the open defects of $C^K$ will become symmetry generators for the 0-form non-invertible $SO(2)$ symmetry of the 4d theory. 

Opening the $C^K$ defect along the $S^2$ cycle, the operators living along what was formerly the $T^2$ direction and now is the open cylinder $\mathcal{C}$, now terminate on the boundary. To ensure gauge invariance of the open defect, we must impose Dirichlet boundary conditions for these operators on $\mathcal{C}$. This means, only the endable surface operators according to the topological boundary conditions in \eqref{8} are allowed to live along this direction, and the integrals over $a,c\in \bR$ charges become sums over integer charges $a,c\in \bZ$,
\begin{equation}
    \begin{split}
        C^K_{open} &=\frac{1}{2(1-\cos \vartheta)}  \sum_{a,c\in \bZ} \int_{b,d \in \bR} e^{\pi i (\frac{\sin \vartheta}{\cos\vartheta -1}) (e\,ab-\frac{e \theta}{2\pi}bc+\frac{1}{e}dc)} e^{\pi i (\frac{1}{e} bc - e \, ad + \frac{e\theta}{2\pi} cd)}\\
        &\times  U_{ae} U_{-e \theta c /2\pi} V_{c/e}(\mathcal{C}) \,U_b V_d(S^2)\,.
    \end{split}
\end{equation}
Notice that the phases have also been affected by the inclusion of the $\theta$ dependent charges introduced in the boundary-ending operators, through the substitution $a \to e\, a - \frac{e\theta}{2\pi} c$. The action of this open defect clearly demonstrates its non-invertibility. For example, acting on $U_x(\mathcal{C})$, the integrals over $b,d$ produce the respective delta functions,
\begin{equation} \label{10}
    \begin{split}
        \delta \left(  a \frac{e\sin \vartheta }{2(\cos \vartheta - 1) } - c \frac{e\theta}{4\pi} \frac{\sin \vartheta}{\cos \vartheta - 1} + c \frac{1}{2e}
        \right)
        \quad , \quad
       \delta \left( -x+ c\frac{\sin \vartheta}{2e (\cos\vartheta - 1) } - a \frac{e}{2}+ c \frac{e\theta}{4\pi } \right)\,.
    \end{split}
\end{equation}
The first delta is only able to be satisfied for $a,c\in \bZ$ when the following rationality condition holds
\begin{equation}\label{9}
    \left(\frac{\theta}{2\pi} - \frac{1}{e^2} \frac{(\cos \vartheta-1)}{\sin \vartheta}\right) \in \bQ  \,.
\end{equation}
If this is not the case, then the action of the defect is zero, essentially projecting out all surface operators, in the sense that they become non-genuine after acting with $C^K_{open}$. The second delta function gives an additional condition on the charge $x$ that must be satisfied for the action on $U_x(\mathcal{C})$ to be nonzero. A similar analysis can be performed when acting on the $V(\mathcal{C})$ operators. The fact that when the rationality condition is satisfied, there exists a subset of operators that remain genuine after acting with $C^K_{open}$ is related to the fact that the boundary Maxwell theory in this case is self-dual under gauging of a discrete subset of the symmetries instead of the entire $U(1)^w$. This projecting out of all or some of the operators is a demonstration of the non-invertible nature of the defect. 

As a specific example, let us consider the defect corresponding to the rotation with $\vartheta= -\pi /2$, i.e. the $S$ transformation. In this case, the delta functions in \eqref{10} simplify to the conditions, 
\begin{equation}
    \begin{split}
        \delta \left(  a e - c \frac{e\theta}{2\pi} + c \frac{1}{e}
        \right)
        \quad , \quad
       \delta \left( -x+ c\frac{1}{2e  } - a \frac{e}{2}+ c \frac{e\theta}{4\pi } \right)\,.
    \end{split}
\end{equation}
Taking the case where the theta angle is zero or an integer multiple of $2\pi$,  the first delta function then implies that $e^2\in \bQ$. This is exactly the rationality condition one would expect in order to construct a self-duality symmetry from composing the $S$ transformation and discrete gaugings. 

If $e^2= \frac{p}{q}$ for some integers $p,\,q$ with $gcd(p,q)=1$, then the second delta function implies that the defect will `throw out' surface operators $U_x(\mathcal{C})$ unless the charge $x$ lies in the overlapping charge lattice of $e\bZ$ and $\bZ/e$, such that $x\in \sqrt{pq}\bZ$. (To see this note that from our boundary conditions in \eqref{10}, $U_x(\mathcal{C})$ is a priori only a sensible boundary ending operator if $x\in e\bZ$.) Notably, if $\tau =i$, $e^2=1$ then all the boundary ending $U$ operators are sent to $V$ operators, no operators are thrown out, and $C^{K=-\pi/2}$ is invertible. Conversely, when $e^2$ and/or $\theta/2\pi$ is irrational, then the defect projects out all operators. We note that these rationality conditions, found using the action of the open defect on operators, match with what is known in the existing literature, in particular with that in \cite{Niro:2022ctq}. 

\section{Defect Fusion} \label{sec5}
In this section, we utilize the parallel fusion of defects to demonstrate various properties of $C^T$, $C^A$, and $C^K$. In particular, we would like to exemplify via fusion that the defects defined on closed manifolds are generators of the $SL(2,\bR)$ symmetry.\footnote{We point out that the ability to define non-Abelian defects via higher form symmetries here is due to the fact that these defects are codimension-one.} Additionally, fusion of the open $C^K$ defect should show the $SO(2)$ nature of these defects as well as their non-invertibility. For completeness, we will consider these fusions from the perspective of summations over surface operators along non-trivial 2-cycles, as well as from the path integral perspective. This also underscores the ease in which certain computations can be done in one perspective over the other.

\subsection{Summation Over Operators Perspective}
Firstly, we tackle the fusion of various defects from the perspective of their mesh of surface operators. We will by no means be exhaustive on detailing the fusion of all possible combinations of our defects, but rather focus on a few examples which we argue are demonstrative of important properties we wish to highlight.

\subsubsection{Invertibility: Closed Defect Fusion}
In the previous sections, we claimed that the $C^T$, $C^A$, and $C^K$ defects are invertible when defined on closed manifolds. Here, we show this explicitly for one particular example, by fusing $C^{A_\lambda}$ with $C^{A_{1/\lambda}}$, demonstrating its invertibility and serving as a check that the fusion is given by the group law for $SL(2,\bR)$. We leave the other examples as an exercise for the interested reader. Using the form of $C^A$ in \eqref{18},
\begin{equation}
    \begin{split}
        C^{A_\lambda} \times C^{A_{1/\lambda}} (T^2 \times S^2) &=  \mathcal{N} \int_{a... \in \bR} e^{2\pi i \Lambda (wz+xy) }  e^{\pi i (xy-wz)} e^{-2\pi i \Lambda (cd+ab)} e^{\pi i (ab-cd)}\\
        &\times  U_w(T^2) V_y(T^2) U_x(S^2) V_z(S^2) \;U_c(T^2) V_b(T^2) U_a(S^2) V_d(S^2)
    \end{split}
\end{equation}
where the normalization factor $\mathcal{N}= \frac{\lambda^2}{(\lambda-1)^4}$ and we have used that $\Lambda= \frac{\lambda+1}{2(\lambda-1)}$ and $-\Lambda= \frac{1/\lambda+1}{2(1/\lambda-1)}$. Moving the $S^2$ dependent operators to the right-hand side to pick up additional unlinking phases given by \eqref{braid}, and fusing like operators,
\begin{equation}
    \begin{split}
        C^{A_\lambda} \times C^{A_{1/\lambda}} (T^2 \times S^2) &=  \mathcal{N} \int_{a... \in \bR} e^{2\pi i \Lambda (wz+xy) }  e^{\pi i (xy-wz)} e^{-2\pi i \Lambda (cd+ab)}e^{\pi i (ab-cd)} e^{2\pi i (bx-cz)}\\
        &\times  U_{w+c}(T^2) V_{y+b}(T^2) U_{x+a}(S^2) V_{z+d}(S^2) \,. 
    \end{split}
\end{equation}
We perform the following redefinition of the summation variables,
\begin{equation}
    w+c = \ell \quad y+b = m \quad x+a = n \quad z+d = p \,.
\end{equation}
This allows us to re-express the above phase factor as,
\begin{equation}
    \begin{split}
        &e^{2\pi i \Lambda (wz+xy) }  e^{\pi i (xy-wz)} e^{-2\pi i \Lambda (cd+ab)}e^{\pi i (ab-cd)} e^{2\pi i (bx-cz)}
        =\\
        &e^{2\pi i \Lambda (\ell p - \ell d -cp +nm -nb -am ) } e^{\pi i (nm+nb-am - \ell p +\ell d -cp )}\,.
    \end{split}
\end{equation}
such that we can integrate over the charges $a,b,c,d$, which altogether give the delta functions with overall normalization,
\begin{equation}
    \frac{(\lambda-1)^4}{\lambda^2} \delta(\ell)\delta(m) \delta(n) \delta(p)\,.
\end{equation}
These delta functions completely trivialize the fused defect, and also cancel the normalization factor (again demonstrating that we chose the correct normalization in our construction). 
\begin{equation}
    C^{A_\lambda} \times C^{A_{1/\lambda}} (T^2 \times S^2)= 1
\end{equation}

\subsubsection{Non-Abelianness: ST Versus TS Defects}
Given that these defects generate the $SL(2,\bR)$ 0-form symmetry of the bulk, we expect to see the non-Abelian nature of these defects from fusion arguments. One simple demonstration of this is the in-equivalence of the oppositely ordered fusion products $C^{T_{\ell=1}} \times C^{K_{\vartheta=-\pi/2}} \neq C^{K_{\vartheta=-\pi/2}} \times C^{T_{\ell=1}} $. In the following, we will show this in-equivalence explicitly, using $C^{T_{\ell=1}} =C^T$ and $C^{K_{\vartheta=-\pi/2}} = C^S$ to lighten the notation. 
\begin{equation}
    \begin{split}
        C^T \times C^S (T^2\times S^2)& = \frac{1}{2}\int_{x,y,a,b,c,d} e^{-2\pi i xy} e^{\pi i (ab+dc)}e^{\pi i (bc-ad)}\\
        &\times U_x(T^2) U_y(S^2) \; U_a(T^2) V_c(T^2) U_b(S^2) V_d(S^2)
    \end{split}
\end{equation}
Following our convention, we move the $S^2$ operators to the right picking up a phase factor.
\begin{equation}
    C^T \times C^S = \frac{1}{2}\int_{x,y,a,b,c,d} e^{-2\pi i xy} e^{\pi i (ab+dc)}e^{\pi i (bc-ad)} e^{2\pi i yc} U_{a+x}(T^2) V_c(T^2) U_{b+y}(S^2) V_d(S^2)
\end{equation}
Defining $a+x=n$ and $b+y=m$, and rewriting the phase factor in terms of these new variables,
\begin{equation}
    C^T \times C^S = \frac{1}{2}\int_{n,m,a,b,c,d} e^{2\pi i (nb-nm+am+mc) } e^{\pi i (dc-ab-ad-bc)} U_{n}(T^2) V_c(T^2) U_{m}(S^2) V_d(S^2)\,.
\end{equation}
This allows us to integrate over $a$,
\begin{equation}
    \int_{\infty}^\infty da\, e^{\pi i\, a(2m-b-d)} = 2\, \delta(b=2m-d)
\end{equation}
such that integrating over $b$ enforces the delta function and the resulting defect is
\begin{equation}
    C^T \times C^S (T^2 \times S^2) =\int_{n,m,c,d \in \bR} e^{2\pi i (nm-nd+dc) } U_{n}(T^2) V_c(T^2) U_{m}(S^2) V_d(S^2)\,.
\end{equation}

Running through the same steps for the oppositely ordered fusion, the only difference with respect to above is that when moving the $S^2$ dependent operators to the right, we incur the unlinking phase $e^{-2\pi i dx}$. The result is given by,
\begin{equation}
    C^S \times C^T(T^2 \times S^2) = \int_{n,m,c,d\in \bR} e^{2\pi i (nm+cm+dc) } U_{n}(T^2) V_c(T^2) U_{m}(S^2) V_d(S^2)\,.
\end{equation}
Clearly, these are in-equivalent, and this non-Abelianess ultimately stems from the ordering/unlinking of operators. 

\subsubsection{SO(2) Fusion of K Defects}
Additionally, we can use the fusion of $C^{K_\vartheta}$ with $C^{K'_{\vartheta'}}$ to demonstrate that they follow a $SO(2)$ group law. That is, fusing the defects corresponding to rotations by $\vartheta$ and $\vartheta'$ the result should also be a rotational transformation by the angle $\vartheta+\vartheta'$. Using the form of the defect in \eqref{11},
\begin{equation}
    \begin{split}
        C^K \times C^{K'} (T^2 \times S^2) &=\mathcal{N} \int_{a,... \in \bR} e^{\pi i \Theta (ab+dc)} e^{\pi i( bc-ad)}e^{\pi i \Theta' (wx+yz)} e^{\pi i (xy-wz)} \\
        &\times U_a(T^2) V_c(T^2) U_b(S^2) V_d(S^2)\; U_w(T^2) V_y(T^2) U_x(S^2) V_z(S^2)\,. 
    \end{split}
\end{equation}
where the normalization factor is $\mathcal{N}=\frac{1}{2(1-\cos \vartheta)} \frac{1}{2(1-\cos \vartheta')}  $ and we have named $\Theta = \frac{\sin \vartheta}{\cos\vartheta -1}$ and $\Theta' = \frac{\sin \vartheta'}{\cos \vartheta'-1}$ to lighten the notation. Moving $S^2$ dependent operators to the right-hand side, we pick up the unlinking phase $e^{2\pi i (by-dw)}$. Fusing like operators motivates the following redefinition of integration variables,
\begin{equation}
    a+w=\ell \quad c+y = m \quad b+x = n \quad d+z=p  \,,
\end{equation}
allowing us to also re-express the phase factor such that the fused defect becomes,
\begin{equation}
    \begin{split}
        C^K \times C^{K'} (T^2 \times S^2) &= \mathcal{N} \int_{a,... \in \bR} e^{\pi i \Theta (ab+dc)}e^{\pi i\Theta' (\ell n - \ell b -an +ab +mp -md-cp+cd)}  \\
        &\times e^{\pi i (bc-ad+nm-nc-bm+bc-\ell p + \ell d +ap -ad +2bm -2bc-2d\ell +2da)} \\
        &\times U_\ell(T^2) V_m(T^2) U_n(S^2) V_p(S^2)\,.
    \end{split}
\end{equation}
Integrating over $a,c$, produces the following Dirac delta functions. (In what follows, we will not be concerned with a detailed treatment of the normalization factor).
\begin{equation}
    \delta\left(b = \frac{\Theta'n -p}{\Theta + \Theta'}\right) \quad , \quad \delta\left(d= \frac{\Theta' p +n }{\Theta + \Theta'}\right)
\end{equation}
The integrals over $b,d$ enforce these delta functions, leaving 
\begin{equation}
    \begin{split}
        C^K \times C^{K'} (T^2 \times S^2) &=  \int_{a,... \in \bR} e^{\pi i \big(\frac{\Theta \Theta'-1}{\Theta + \Theta'}\big) (\ell n +mp)} e^{\pi i (nm-p\ell )}  U_\ell V_m(T^2)\, U_nV_p(S^2)\,.
    \end{split}
\end{equation}
Using trigonometric identities and comparing to the definition of the defect in \eqref{11},
\begin{equation} \label{12}
    \begin{split}
        C^K \times C^{K'} (T^2 \times S^2) &=  \int_{a,... \in \bR} e^{\pi i \big(\frac{sin(\vartheta +\vartheta')}{cos(\vartheta+\vartheta')-1}\big)(\ell n +mp)} e^{\pi i (nm-p\ell )}  U_\ell V_m(T^2) \,U_nV_p(S^2)\\
        &= C^{K_{\vartheta + \vartheta'}} (T^2 \times S^2) \,.
    \end{split}
\end{equation}
The defect fuses as expected for $SO(2)$.

\subsection{Path Integral Perspective}
Throughout this note so far, we have focused on defining our defects as a mesh of $U(\Gamma),V(\Gamma^{\prime})$ operators along non-trivial cycles. Now, we wish to pivot to describe this higher gauging from the path integral perspective. To do so, we use Poincar\'e duality to re-express the non-trivial cycles that we are summing over as gauge fields, $\phi = PD_\Sigma(\Gamma)$, $\psi =PD_\Sigma (\Gamma')$, coupled to the conserved current(s) associated with the symmetries being gauged. It is important that these gauge fields $\phi,\,\psi$ are flat as this ensures that the defects are topological, see Appendix \ref{app a}, a condition which is in practice enforced by a Langrange multiplier in the defect action. 

In what follows, we first define each of our condensation defects via their actions on the codimension-one surface $\Sigma$, and then move on to describe their fusion. 

\subsubsection{The Defect Actions}
Using Poincar\'e duality it is straightforward to translate between the summation over operators and path integral perspectives, such that we find the defect action for $C^T$, 
\begin{equation}
    \mathcal{L}^T(\Sigma) = \frac{i}{2\pi } \int_\Sigma (b \phi + \phi d \chi) - \frac{i \ell^{-1} }{4\pi} \int_{\Sigma} \phi \wedge \phi \,. 
\end{equation}
Here $\phi$ is the 2-form gauge field responsible for gauging the $\bR^e$ symmetry, whose conserved current is $b$, and $\chi$ is a 1-form Lagrange multiplier enforcing the flatness of $\phi$. Both $\phi,\,\chi$ are real-valued fields. For gauge invariance, we require that $\chi$ has the gauge transformation in terms of the gauge parameter of $b$,
\begin{equation}
    \chi \to \chi- \lambda_b\,.
\end{equation}
Using the flatness of the gauge fields $\phi,\,b,\,c$, coming from the equations of motion of $\chi$ and the bulk equations of motion respectively, this defect action only remains gauge invariant under $\phi \to \phi + d \lambda_\phi$ when $\Sigma$ does not have a boundary. 
When $\partial \Sigma \neq 0$ we must include additional boundary terms to ensure gauge invariance. Introducing the 1-form field $\sigma$ with gauge transformations $\sigma \to \sigma - \lambda_\phi$, the necessary boundary terms are 
\begin{equation}
    \mathcal{L}^T + \frac{i}{2\pi } \int_{\partial \Sigma}( b\sigma + \sigma d \chi )- \frac{i \ell^{-1}}{4\pi} \int_{\partial \Sigma} (\sigma d \sigma +2 \phi \sigma ) \,.
\end{equation}

Moving on, the defect action for $C^A$ is given by,
\begin{equation}
    \mathcal{L}^A = \frac{i}{2\pi} \int_{\Sigma} (\phi  b+\phi'c+ \phi d\chi+ \phi'd\chi') + \frac{i}{4\pi}  \frac{1+\lambda}{2(\lambda-1)} \int_{\Sigma} \phi\wedge \phi' \,,
\end{equation}
where we have introduced the Lagrange multipliers $\chi,\; \chi'$ with gauge transformations $\chi \to \chi - \lambda_b$ and $\chi' \to \chi' - \lambda_c$. If the $\Sigma$ has a boundary, the defect action is supplemented with the following boundary terms, which ensure gauge invariance under the gauge transformations of $\phi$ and $\phi'$.
\begin{equation}
    \mathcal{L}^A + \frac{i}{2\pi} \int_{\partial \Sigma} (\sigma b + \sigma' c +\sigma d \chi + \sigma' d \chi' ) + \frac{i}{4\pi}  \frac{1+\lambda}{2(\lambda-1)} \int_{\partial \Sigma} (\sigma \phi' + \phi \sigma' + \sigma d\sigma')
\end{equation}
Here we have introduced the 1-forms $\sigma$ and $\sigma'$ with gauge transformations $\sigma^{(\prime)} \to \sigma^{(\prime)} - \lambda_{\phi^{(\prime)}}$. 

Lastly, the defect action for the rotational transformations $C^K$,
\begin{equation}
    \mathcal{L}^K = \frac{i}{2\pi } \int_{\Sigma}( \phi b+\phi'c + \phi d \chi + \phi'd\chi') +\frac{i \sin \vartheta}{8\pi(\cos \vartheta-1)} \int_\Sigma( \phi \wedge \phi +  \phi' \wedge \phi')\,. 
\end{equation}
Again to ensure gauge invariance in case of boundary, we must add boundary terms, 
\begin{equation}\label{13}
    \mathcal{L}^K + \frac{i}{2\pi } \int_{\partial \Sigma}( \sigma b + \sigma' c + \sigma d \chi + \sigma' d \chi') + \frac{i \sin \vartheta}{8\pi(\cos \vartheta - 1)} \int_{\partial \Sigma} (\sigma d \sigma+2\phi \sigma + \sigma' d \sigma' +2\phi' \sigma')\,,
\end{equation}
where we have once again introduced the 1-forms $\sigma,\,\sigma'$ that transform under the same gauge parameters as $\phi,\phi'$ respectively, $\sigma^{(\prime)} \to \sigma^{(\prime)} - \lambda_{\phi^{(\prime)}}$.

\subsubsection{NonInvertibility: Fusion of Open Defects}
The path integral perspective is particularly well-suited for computing the fusion of open defects. As discussed previously, we expect only the open $C^K$ defect to be a generator of the generically non-invertible self-duality symmetry of a given 4d theory. This non-invertibility should be apparent from the fusion of the opened $C^K$ defects. In \eqref{12}, we found that the defect $C^{K_\vartheta}$ defined on a manifold without boundary has the inverse $C^{K_{-\vartheta}}$. Here, we will show that once the defects are opened, the inverse no longer exists, and the defects are non-invertible.

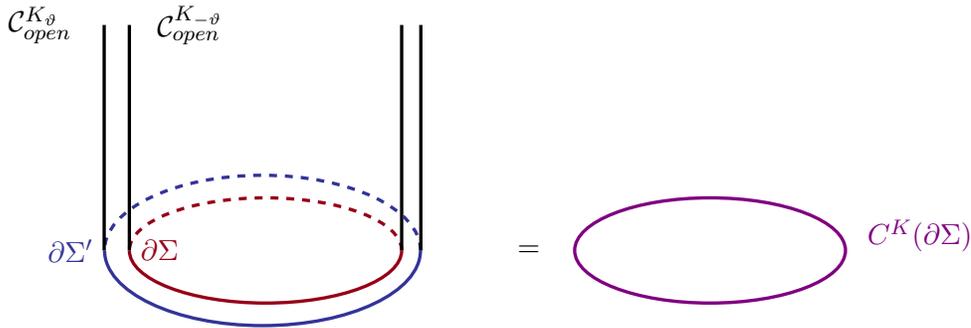
\begin{figure}[ht]
\centering
    \begin{tikzpicture}
    \draw [dashed,carmine,line width = 1.25] (0,2) +(0:1.81 and 0.7) arc [start angle = 0, end angle = 180, x radius = 1.81, y radius = 0.7]node[right]{$\partial \Sigma$};
    
\draw [carmine,line width = 1.25] (0,2) +(0:1.81 and 0.7) arc [start angle = 0, end angle = -180, x radius = 1.81, y radius = 0.7];
\draw[line width = 1.25] (-1.81, 2) to (-1.81, 5);
\draw[line width = 1.25] (1.81, 2) to (1.81, 5);
\node[] at (-1.,5) {$\mathcal{C}_{open}^{K_{-\vartheta}}$};

\draw [dashed,blueflag,line width = 1.25] (0.25,2.) +(0:1.81 and 0.7) arc [start angle = 0, end angle = 180, x radius = 2.1, y radius = 1.]node[left]{$\partial \Sigma^\prime$};
\draw [blueflag,line width = 1.25] (0.25,2.) +(0:1.81 and 0.7) arc [start angle = 0, end angle = -180, x radius = 2.1, y radius = 1.];
\draw [line width = 1.25](-2.145, 2) to (-2.145, 5);
\draw[line width = 1.25] (2.065, 2) to (2.065, 5);
\node[] at (-3.,5) {$\mathcal{C}_{open}^{K_\vartheta}$};

\node[] at (3.5,2){$=$};

\begin{scope}[shift={(6.2,.2)}]
\draw [violet,line width = 1.25] (-2.1,2.5) +(0:1.81 and 0.7) arc [start angle = -270, end angle = 90, x radius = 1.8, y radius = 0.7];
\node[violet] at (2.5,2) {$C^{K}(\partial \Sigma)$};
\end{scope}

\end{tikzpicture}
\hspace{1cm}
    \caption{Schematically, the fusion of open defects $C^{K_{\vartheta}}(\Sigma)$ with $C^{K_{-\vartheta}}(\Sigma')$. The `bulk part' of the defects fuses to the identity, leaving only a boundary contribution.}
    \label{fig:fusion of open K defects}
\end{figure}

The fusion of $C^{K_\vartheta} \times C^{K_{-\vartheta}}$ is given by adding together their respective defect actions $\mathcal{L}^{K_\vartheta}[\phi, \phi^\prime; \Sigma]$ and $\mathcal{L}^{K_{-\vartheta}}[\psi, \psi'; \Sigma']$ (see equation \eqref{13}), as well as including the bulk contribution of the slice of spacetime between the defects \cite{Antinucci:2022vyk, Argurio:2024ewp}. 
To compute the bulk contribution, notice that the higher gauging defining the defects changes the equations of motion to,
\begin{equation}
    \begin{split}
        db &= \phi' \delta(\Sigma) + \psi' \delta(\Sigma') \\
        dc &=- \phi \delta(\Sigma) - \psi \delta(\Sigma') \,,\\
    \end{split}
\end{equation}
where the delta functions are localized on the respective defects. Since all of the gauge fields involved are flat on-shell we can integrate both sides of these equations,
\begin{equation}
    \begin{split}
        b &= \phi' \Theta(\Sigma) + \psi' \Theta(\Sigma') \\
        c &=- \phi \Theta(\Sigma) - \psi \Theta(\Sigma') \,,\\
    \end{split}
\end{equation}
where $\Theta(\Sigma^{(\prime)})=\Theta(x-x_{\Sigma^{(\prime)}})$ is the Heaviside step function and $x$ is the coordinate measuring the separation between the surfaces $\Sigma, \, \Sigma'$. Substituting the above into the (symmetrized) bulk action as well as the defect actions, the bulk contribution depends on the ordering of $\Sigma$ and $\Sigma'$. If $\Sigma'$ is to the left of $\Sigma$ the contribution is,
\begin{equation}
    \frac{1}{2} \phi \psi' - \frac{1}{2}\phi' \psi \,;
\end{equation}
alternatively, if $\Sigma'$ is to the right of $\Sigma$ it is,
\begin{equation} \label{888}
    - \frac{1}{2} \phi \psi' + \frac{1}{2}\phi' \psi\,.
\end{equation}
Choosing the case that $\Sigma'$ is to the left of $\Sigma$, and putting this altogether,
\begin{equation}
    \begin{split}
          C^{K_{-\vartheta}} \times C^{K_\vartheta} &= \frac{i}{2\pi } \int_{\Sigma}( (\phi+\psi) b+(\phi'+\psi')c + \phi d \chi + \phi'd\chi' + \psi d \zeta + \psi' d\zeta'+ \frac{1}{2} (\phi \psi' - \phi' \psi )) \\
        &+ \frac{i \sin \vartheta}{8\pi(\cos \vartheta - 1)}  \int_\Sigma( \phi  \phi + \phi'\phi' - \psi \psi -\psi'\psi') \\
        &+ \frac{i}{2\pi } \int_{\partial \Sigma} ((\sigma+\rho) b + (\sigma'+\rho') c + \sigma d \chi + \sigma' d \chi' + \rho d \zeta + \rho' d \zeta') \\
        &+ \frac{i \sin \vartheta}{8\pi(\cos \vartheta - 1)}  \int_{\partial \Sigma} (\sigma d \sigma +2\phi \sigma +\sigma' d\sigma' + 2\phi'\sigma' -\rho d \rho - 2\psi \rho - \rho' d \rho' -2\psi' \rho')\,.
    \end{split}
\end{equation}
Making the field redefinitions,
\begin{equation}
    \begin{aligned}
        \Tilde{\phi} &= \phi + \psi \qquad \hat{\phi} &= \phi - \psi \qquad \Tilde{\chi} &= \chi + \zeta \\
        \hat{\chi} &= \chi - \zeta \qquad 
        \Tilde{\sigma} &= \sigma + \rho \qquad \hat{\sigma} &= \sigma - \rho \,,\\
    \end{aligned}
\end{equation}
and analogously for the primed fields, the fusion becomes, 
\begin{equation} \label{91}
    \begin{split}
        C^{K_{-\vartheta}} \times C^{K_\vartheta} &= \frac{i}{2\pi } \int_{\Sigma}( \Tilde{\phi} b+\Tilde{\phi}'c + \frac{1}{2}(\Tilde{\phi} d \Tilde{\chi} + \hat{\phi} d \hat{\chi} + \Tilde{\phi}' d \Tilde{\chi}' + \hat{\phi}' d \hat{\chi}') +\frac{1}{4}(\Tilde{\phi}' \hat{\phi} - \Tilde{\phi}\hat{\phi}') ) \\
        &+ \frac{i \sin \vartheta}{8\pi(\cos \vartheta - 1)} \int_\Sigma( \Tilde{\phi} \hat{\phi} + \Tilde{\phi}' \hat{\phi}') \\
        &+ \frac{i}{2\pi } \int_{\partial \Sigma} (\Tilde{\sigma} b + \Tilde{\sigma}' c + \frac{1}{2}(\Tilde{\sigma} d \Tilde{\chi} + \hat{\sigma} d\hat{\chi}+ \Tilde{\sigma}' d \Tilde{\chi}' + \hat{\sigma}' d\hat{\chi}')) \\
        &+ \frac{i \sin \vartheta}{8\pi(\cos \vartheta - 1)}  \int_{\partial \Sigma} (\Tilde{\sigma} d \hat{\sigma} +\Tilde{\phi}\hat{\sigma} + \hat{\phi} \Tilde{\sigma }+ \Tilde{\sigma}' d \hat{\sigma}' + \Tilde{\phi}' \hat{\sigma}' + \hat{\phi}' \Tilde{\sigma}' )\,.
    \end{split}
\end{equation}
The path integrals over $\hat{\phi}$ and $\hat{\phi}'$ on $\Sigma$ impose that $\Tilde{\phi}$ and $\Tilde{\phi}'$ are linear combinations of $d\hat{\chi}$ and $d\hat{\chi}'$,
\begin{equation}
    \begin{split}
        \Tilde{\phi} &= - \frac{2}{(\Theta + \frac{1}{\Theta})}d\hat{\chi} + \frac{2}{(\Theta^2+1)} d\hat{\chi}' = d\eta \\
        \Tilde{\phi}' &= - \frac{2}{(\Theta^2+1)} d\hat{\chi} - \frac{2}{(\Theta +\frac{1}{\Theta}) }d\hat{\chi}' = d\eta'\\
    \end{split}
\end{equation}
where we have employed the notation that $\Theta= \frac{\sin \vartheta}{\cos \vartheta -1}$. Path integrating over $\Tilde{\phi}$ and $\Tilde{\phi}'$ enforces that they are proportional to exact forms, which we have named $d\eta,\, d\eta'$ respectively. Using the bulk equations of motion, this acts to trivialize all of the terms on $\Sigma$ and induce additional boundary terms on $\partial \Sigma$.
\begin{equation}
        \frac{i}{2\pi} \int_{\partial \Sigma}( \eta b +\eta' c + \frac{1}{2}(\eta d\Tilde{\chi} +\eta' d\Tilde{\chi}'))  
\end{equation}
Meanwhile, the path integrals over $\hat{\phi}$ and $\hat{\phi}'$ on $\partial \Sigma$ act as delta functions that set $\Tilde{\sigma}=0$ and $\Tilde{\sigma}'=0$ respectively. Furthermore, the path integral over $\Tilde{\phi}, \Tilde{\phi}'$ on $\partial \Sigma$ implies $\hat{\sigma},\,\hat{\sigma}'=0$. Combining this altogether, and rescaling the $\Tilde{\chi}, \, \Tilde{\chi}'$ fields (allowed since they are $\bR$-valued), the result of the fusion contains only the boundary terms,
\begin{equation}
    C^{K_{-\vartheta}} \times C^{K_\vartheta} = \frac{i}{2\pi} \int_{\partial \Sigma}( \eta b +\eta' c + \eta d\Tilde{\chi} +\eta' d\Tilde{\chi}')\,.
\end{equation}
Thus, the fusion result is a three-dimensional condensation defect from gauging the same $\bR\times \bR$ symmetry with flat connections. After slab compactification, imposing the topological and physical boundary conditions, in the 4d theory this gauging corresponds to gauging the $U(1)^e \times U(1)^m$ or a discrete subgroup of this 1-form symmetry depending on the specific choice of $e,\, \theta$. In this way, the bulk condensation defect reproduces all of the self-duality symmetry defects of 4d Maxwell. 

Note that if we redid the computation, starting instead with $\Sigma'$ positioned to the right of $\Sigma$, the result is the again a defect corresponding to gauging the $\bR\times \bR$ symmetry on the 3d boundary. To see this, note that taking the bulk contribution in this case to be \eqref{888} only changes the term $\frac{1}{4}(\Tilde{\phi}' \hat{\phi} - \Tilde{\phi} \hat{\phi}')$ in the fusion product in \eqref{91} to $\frac{1}{4}(\Tilde{\phi} \hat{\phi}' - \Tilde{\phi}'\hat{\phi})$. The path integrals over $\hat{\phi}$ and $\hat{\phi}'$ on $\Sigma$ now impose that
\begin{equation}
    \begin{split}
        \Tilde{\phi} &= - \frac{2}{(\Theta + \frac{1}{\Theta})}d\hat{\chi} - \frac{2}{(\Theta^2+1)} d\hat{\chi}'  \\
        \Tilde{\phi}' &= - \frac{2}{(\Theta^2+1)} d\hat{\chi} + \frac{2}{(\Theta +\frac{1}{\Theta}) }d\hat{\chi}' \,.\\
    \end{split}
\end{equation}
Naming these new linear combinations of $\hat{\chi}, \, \hat{\chi}'$ as $\Tilde{\phi}= d\xi$ and $\Tilde{\phi}' = d\xi'$, due to the bulk equation of motion $db=dc=0$ this again acts to trivialize all of the terms on $\Sigma$. Running through the same steps as before we find that the fusion becomes,
\begin{equation}
    C^{K_\vartheta} \times C^{K_{-\vartheta}} = \frac{i}{2\pi} \int_{\partial \Sigma}( \xi b +\xi' c + \xi d\Tilde{\chi} +\xi' d\Tilde{\chi}')
\end{equation}
which again is interpreted as gauging the $\bR\times \bR$ symmetry with flat connections on $\partial \Sigma$.

\section*{Acknowledgments}
The author is grateful to Adrien Arbalestrier, Riccardo Argurio, Giovanni Galati, Ondrej Hulik, and Valdo Tatitscheff for insightful discussions. E.P. thanks Riccardo Argurio and Giovanni Galati for their feedback on a preliminary version of this work. The research of E.P. is funded by an ARC advanced project.

\appendix
\section{Gauging and Topological Defects} \label{app a}
In the main text, we stated that higher-gauging, that is, gauging on submanifolds of the entire spacetime that the theory is defined on, does not change the bulk theory but rather defines a defect within that theory. This defect will only be topological if the gauge field that is performing the said gauging is flat. 

Consider a generic theory defined by the action $S$ living on the manifold $M_d$ with a p-form symmetry given by the conserved current $\star J$. To higher gauge this symmetry on the submanifold $\Sigma_q \subset M_d$ (for some $q>d-p-1$) we couple to the $(p+q-d+1)$-form gauge field $A$ and make $A$ dynamical.\footnote{Depending on the degree of the forms involved, the gauging here could be done with the addition of discrete torsion. This is from adding a term $\propto A \wedge A$ to the defect action. Translating to the summation over operators perspective, this corresponds to the addition of a phase factor in the defect.} The path integral becomes,
\begin{equation}
    \mathcal{Z} = \int\mathcal{D}[\phi,A] e^{iS}e^{i\int_\Sigma A \wedge \star_d J}\,.
\end{equation}
The term $e^{i\int_\Sigma A \wedge \star_d J}$ defines a defect on $\Sigma_q$, which we name $C(\Sigma_q)$ for simplicity in the following argument. We claim that such a defect is topological if and only if $dA=0$. If the defect is topological, it is invariant under smooth deformations of $\Sigma_q$ that do not cross over any other operator insertions. Let us consider such a deformation sending $\Sigma \to \Sigma'$. $\Sigma, \, \Sigma'$ can be thought of as boundaries of a $q+1$-dimensional manifold $\partial X_{q+1}= \Sigma_q'- \Sigma_q$. The corresponding change in the defect becomes,
\begin{equation}
    \begin{split}
        C(\Sigma_q') &= C(\Sigma_q + \delta X_{q+1}) = \exp\left(i\int_{\Sigma_q} A \wedge \star_d J+ i\int_{\partial X_{q+1}} A \wedge \star_d J\right)  \\
        &= C(\Sigma_q) \; \exp\left(i\int_{X_{q+1}} d(A \wedge \star_d J)\right) \,,
    \end{split}
\end{equation}
where in the last line we have used Stokes' theorem. We see that $C(\Sigma_q) = C(\Sigma'_q)$ if $dA=0$ ($d\star J=0$ due to the conservation equation). However, that the gauge field has flat connections, $dA=0$, is not a priori given. When gauging discrete symmetries, it is always true that $dA=0$, meaning higher gauging of discrete symmetries always corresponds to topological defects.\footnote{For gauging a discrete symmetry in the continuous field formalism, this amounts to adding the additional term to the action on $\Sigma$ which is in charge of `Higgsing' the symmetry down to $\bZ_N$, $\propto N \int_\Sigma A d\psi$. This makes it clear that $dA=0$ in the discrete case.} When gauging continuous symmetries, this is not necessarily the case, and the defect is only topological if we choose to impose the flatness condition. 

We now assume that the condition $dA=0$ holds, (in practice this is implemented by a Lagrange multiplier). Given the flatness of $A$, the gauge field is completely characterized by its holonomies. Therefore, we can decompose the path integral over the $A$ into a sum over representatives of the cohomology classes of the manifold $\Sigma_q$, $A_w \in H^{p+q-d+1}(\Sigma_q)$.
\begin{equation}
    \mathcal{Z} = \sum_{A_w \in H^{p+q-d+1}(\Sigma_q)} \int\mathcal{D}[\phi] e^{iS}e^{i\int_\Sigma A_w \wedge \star_d J}
\end{equation}
Applying Poincar\'e duality $PD_q(A) =\Gamma \in H_{d-p-1}(\Sigma_q)$,
\begin{equation}
    \mathcal{Z} =  \int\mathcal{D}[\phi] e^{iS} 
  \sum_{\Gamma \in H_{d-p-1}(\Sigma_q)} e^{i\int_\Gamma \star_d J} \,,
\end{equation}
which is exactly how we defined the condensation defects from the summation over operators. Thus, higher gauging when done with flat connections is equivalent to the insertion of topological defects in the path integral.

\bibliography{bib}
    \bibliographystyle{ytphys}
\end{document}